\documentclass[preprint]{aastex}

\usepackage{psfig}

\def\ifundefined#1{\expandafter\ifx\csname#1\endcsname\relax}

\def\la{\mathrel{\hbox{\rlap{\hbox{\lower4pt\hbox{$\sim$}}}\hbox{$<$}}}}
\def\ga{\mathrel{\hbox{\rlap{\hbox{\lower4pt\hbox{$\sim$}}}\hbox{$>$}}}}

\newcommand{\be}{\begin{eqnarray}}
\newcommand{\ee}{\end{eqnarray}}

\ifundefined{ensuremath}\def\ensuremath#1{\relax\ifmmode{#1}}
\else${#1}$\fi\else\relax\fi
\ifundefined{nuc}\def\nuc#1#2{\relax\ifmmode{}^{#1}{\protect\text{#2}}
\else${}^{#1}$#2\fi}\else\relax\fi

\newcommand{\kmps}{km~s$^{-1}$}

\newcommand{\msol}{\ensuremath{{\textrm{M}_\odot}}}

\def\ang{\hbox{\AA}}
\def\Tmod{\ensuremath{T_{\textrm{model}}}}

\def\tstd{\ensuremath{\tau_{\textrm{std}}}}

\def\alog#1{\times 10^{#1}}

\begin{document}

\bibliographystyle{apj}

\title{Spectral Models of the Type Ic SN 1994I in M51}

\author{E.~Baron, David Branch}
\affil{Dept. of Physics and Astronomy, University of
Oklahoma, 440 W.  Brooks, Rm 131, Norman, OK
73019-0225\email{baron@mail.nhn.ou.edu,branch@mail.nhn.ou.edu}}

\author{Peter H.~Hauschildt}
\affil{Dept. of Physics and Astronomy \& Center for Simulational Physics, 
University of Georgia, Athens, GA 30602-2451}
\email{yeti@hal.physast.uga.edu}

\author{Alexei V.~Filippenko}
\affil{Dept. of Astronomy, University of California, Berkeley, CA
94720-3411}
\email{alex@astro.berkeley.edu}

\author{and}

\author{R.~P.~Kirshner}
\affil{CfA, 60 Garden St., Cambridge, MA 02138}
\email{kirshner@cfa.harvard.edu}

\begin{abstract}
We present detailed non-local thermodynamic equilibrium (NLTE)
synthetic spectra for comparison with a time series of 
observed optical spectra of the Type Ic supernova 1994I which occurred
in M51. With the exceptions of Si~I and S~I, we treat the important
species in the formation of the 
spectrum in full NLTE. We present  results for both a hydrodynamic
model that has been fit to the light curve and  for an illustrative
custom crafted model that is more massive.
Both models give reasonable fits to the overall observed spectra;
however, neither is able to reproduce all the observed features. Some
conspicuous observed features are absent and some predicted features
are unobserved. 
No
model that we have explored is able to satisfactorily reproduce the 
observed infrared feature near 1~$\mu m$ on April 15, 1994 (+7d),
which has been 
attributed to the triplet He~I $\lambda10830$ transition.
The low-mass
hydrodynamic model produces an infrared
feature with a blend of He~I, C~I, O~I, and Si~I--II lines, but it
predicts a strong unobserved absorption feature near 6100~\ang\ due to
Fe~III, and the observed feature just blueward of 6000~\ang\
most likely due to Na D is not reproduced.
The more massive model
does a better job of reproducing the 
observed infrared lineshape, but also predicts the
unobserved feature near 6100~\ang. The early-time 
spectrum of the low-mass model is far too blue; thus, a more massive
model may be slightly favored. Since the predicted infrared
feature is produced by a blend of so many elements and there is no
overwhelming evidence for other helium features such as 
$\lambda5876$, it may be premature to conclude that SNe~Ic
unambiguously contain helium. Thus, we conclude that pure C+O cores
are still viable progenitors for SNe~Ic.
\end{abstract}


\section{Introduction}
Type I supernovae are classified on the basis of their spectra. SNe Ia
have strong Si II $\lambda6355$ lines in their spectra where the
absorption minimum is typically shifted to around 6100~\ang\ near
maximum light, April 8, 1994 \citep{rich94i}, and their late-time
spectra are dominated by iron peak 
elements.  SNe Ib have weak or absent Si~II lines, but strong optical
He~I $\lambda5876$ and $\lambda 6678$ at maximum light, with late-time
spectra dominated by lines of [O~I] and [Ca~II]. SNe Ic appear very
similar to SNe Ib at 
late times, but they lack the strong He I lines at early times.

Supernova 1994I, which occurred in the nearby galaxy M51, is undoubtedly
the best studied Type Ic supernova to date
\citep{fil94i,b94i1,whsn94i95,iwsn94i94,nomnat94i,wooseastasi97,rich94i,wlw93,clocc94i96,millard94i99}.
As originally classified \citep{hw90,wh90}, SNe Ic were thought to
show a complete absence of He~I lines. However, \cite{fil94i}
 identified He~I $\lambda 10830$ in the spectrum of SN 1994I
and \cite{clocc94i96} have inferred the presence of weak He~I
$\lambda5876$ and $\lambda 6678$ lines in SN 1994I as well as in SN Ic
1987M.

The apparent absence of helium  in the spectra of the first few
observed SNe~Ic led to
the inference that the progenitor star was likely to be a bare (or nearly
bare) C+O core of a massive star, which explodes due to the collapse
of the iron core to form a neutron star \citep{wh90,hw90}. The helium
envelope of the massive star could be stripped from the star either
through extensive mass loss in a Wolf-Rayet phase, via interaction
with a binary companion, or a combination of the two
\citep{wh90,hw90,fps90,pod92,wlw93,nomnat94i}. On the basis of the
rate of decline of the light curve, Nomoto and collaborators suggested
that the difference between SNe Ib and SNe~Ic was due to the mass and
degree of mixing, with SNe~Ic being both less massive and more
mixed \citep{nfs90,nskhm91,yn91,hmns91,jbfn91,bnf91}. Since helium is
the diagnostic for SNe~Ib, it has been suggested that SNe~Ib
should in fact be the more thoroughly mixed ones
\citep{bar92,wooseastasi97}.

\section{Model Construction}

Hydrodynamic models of SN~1994I have been constructed to fit the light
curve for single star progenitors \citep{wlw93,wooseastasi97} and for
binary progenitors \citep{iwsn94i94}. We use the CO21 model of
\citet{iwsn94i94} (total ejected mass = 0.9~\msol, total deposited 
energy = $1\alog{51}$~ergs, and a parameterized model
constructed by us to 
represent the more massive core favored by Woosley and collaborators
\citep{wlw93,wooseastasi97}.  The parameterized model is constructed
by specifying a density profile, taken to be $\rho \propto r^{-N}$,
with $N=8$, a model temperature $\Tmod=6800$~K, the reference radius
$R_0$, which is the radius where the continuum optical depth in
extinction at 5000~\ang\ (\tstd) is unity, the expansion velocity,
$v_0 = 6000$~\kmps, at the reference radius, which is determined by
fitting the width and position of the observed spectral features (and
using homologous expansion, $R_0 = v_0\,t$ where $t$ is the time since
explosion and we have used an explosion date of Mar 30 in these
calculations, 9 days before B max), and finally the elemental
abundances. The total ejected mass of the model is 2.4~\msol, with a
kinetic energy of $1.3 \alog{51}$~ergs. The
total nickel mass is about 0.1~\msol; however, since we use only a
local deposition function, no strong physical significance should be
attached to this value and it is only indicative (this is not the case
for the hydrodynamical models where the deposition function is
calculated accurately). Setting the elemental abundances leads in
principle to 1950 free parameters, since we include 39 elements and we
must specify the relative number fraction of each of these elements in
each of the 50 radial zones used in the calculations. In order to
minimize the number of abundance parameters we follow our previous
work \citep{b94i1} and begin with a solar abundance \citep{ag89} and
then ``burn'' all the hydrogen to helium. Next we ``burn'' some helium
to carbon and oxygen with a final number ratio He/O, and a production
ratio by number of carbon to oxygen (C/O). Finally we allow metals
heavier than oxygen to be scaled by a factor $Z$ (taken equal to the
solar value in the model we present here, although we have studied the
effects of varying the model content by up to 3 times the solar value,
we find our results to be insensitive to the metallicity in this
range). While in our 
previous models of SN~1994I \citep{b94i1} the composition was uniform
throughout the model, in these calculations that assumption is
relaxed. The hydrodynamic models are fully non-uniform in composition.
In the parameterized models we consider only 2 regions which consist of
differing composition; a primarily helium composition of total mass $M
= 0.06$~\msol\ (for the model
presented here He/O = 10), the remaining portion was primarily C/O
composition (for the 
model presented here He/O = 1, so helium was somewhat mixed throughout
the model). A final parameter of the model was the total
mass of the ejecta. We note that while $\Tmod$ is a parameter used to
fit the spectral shape, if the abundances, velocities, and density
structure are in significantly error no choice of $\Tmod$ will lead to
a reasonable fit to the overall shape of the spectrum.

While we present here the results of our single ``best'' model we base
our conclusions on our total parameter study which for April 15 alone
consisted of 33 different compositions, 5 different total masses and a
range of deposition functions (or degree or mixing of Ni) that spanned
4 orders of magnitude. The parameter study included allowing for 3
differing compositions (a nearly pure helium layer, a mixed C/O + He
layer, and a nearly pure C/O layer). 

Our choice of a local deposition function implies that the nickel is
distributed throughout the atmosphere with a constant mass fraction
and the non-thermal ionization rates are then calculated assuming that
all of the gamma-ray energy instantaneously produced is deposited
locally. Of course radioactive decay of the nickel is accounted
for. Although this is a somewhat crude approximation, it is expedient and
by varying the assumed nickel mass fraction we can simply study the
effects of mixing. At high optical depths where the matter is in LTE,
the assumed deposition function is irrelevant (our results show that
we recover the LTE populations exactly) and thus we really can probe
the effects of nickel mixing in the line forming region with this
simple approximation.

For the hydrodynamical models our procedure was the same as described
in \cite{nughydro97}. The results from the hydrodynamical calculations
were expanded homologously, radioactive decay was included when
adjusting the compositions, and the gamma-ray deposition function was
calculated assuming pure absorptive gamma-ray opacity and local
deposition of positrons. Although \citet{milne99} find that local
deposition of positrons likely overestimates the energy deposition of
positrons at late times in Type Ia supernovae, this assumption should
have negligible effects on the early times we explore here.

\begin{figure*}
\begin{center}
\leavevmode
\psfig{file=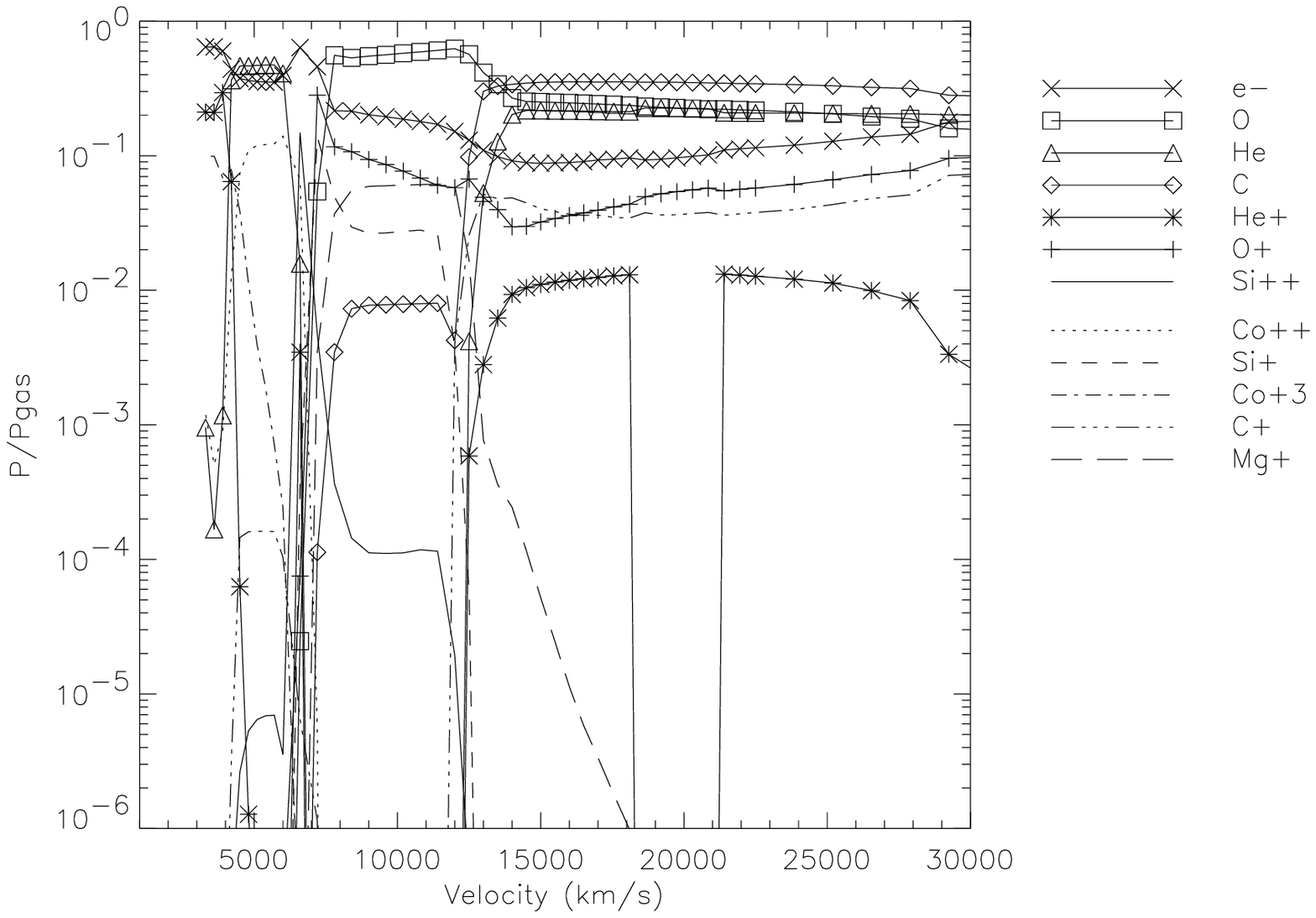}
\caption{\label{abundco21_3463}
 The partial pressures of the most
prevalent species in the CO21 model  as a function of velocity  for the 
April~15 (+7d) epoch. (The synthetic spectrum produced by this model is
shown in 
Figure~\protect\ref{apr15_calc3463}.)}
\end{center}
\end{figure*}

Figures~\ref{abundco21_3463} and \ref{abundfid3_2987} present the most
important species in the two models (CO21 and our more massive
parameterized model for the April 15 [+7d] epoch) as a function of
velocity.  Figure~\ref{vels} displays the velocity as a function of
\tstd\ (the total continuum extinction optical depth at 5000~\ang) in
order to aid in the interpretation of the abundance figures. 
These ions (O~I-II, He~I-II, C~I-II, Si~II-III, Mg~II), are
expected  and support the line
identifications herein and in \citet{millard94i99}. 

The synthetic spectra predicted by the constructed models were then
computed using the generalized stellar atmosphere code {\tt PHOENIX}
\citep[cf.][and references therein]{hbjcam99}.
Table~\ref{nltetab} lists the ions and number of levels/transitions
that we treat in full NLTE in these calculations. {\tt PHOENIX} is
capable of treating a larger number of species in NLTE, but we have
chosen the most important species for these conditions. Even with this
subset of NLTE 
species each model calculation takes about 1 day  using 5
thin-2 nodes of an IBM SP2 parallel supercomputer.

\section{Results}

\begin{figure*}
\begin{center}
\leavevmode
\psfig{file=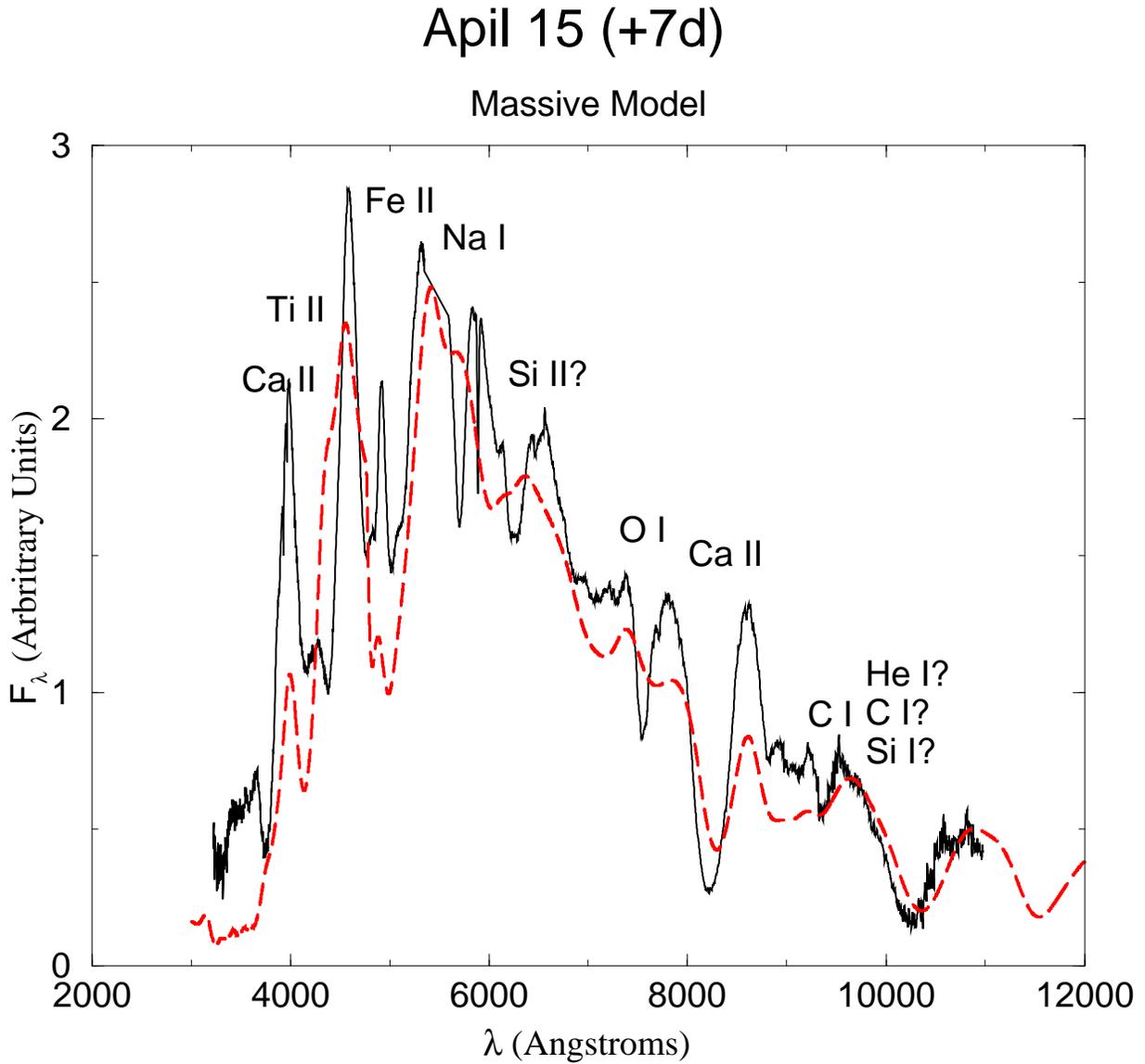}
\caption{\label{apr15_calc2987} The synthetic spectrum of our massive star
model (dashed curve) is compared to the  spectrum taken at Lick Observatory
on April 15, 1994 (+7d) \protect\citep{fil94i}.} 
\end{center}
\end{figure*}

Figure~\ref{apr15_calc3463} displays our synthetic spectrum for the
CO21 model of \cite{iwsn94i94}, compared to the spectrum taken at Lick
Observatory on April 15, 1994 (+7d). The observed spectrum has been dereddened
using the reddening law of \cite{card89} and assuming E(B-V)
=0.45~mag. Since the supernova shows very strong interstellar Na
absorption lines and occurred in a dust lane in M51, the reddening is
very uncertain. Values of E(B-V) in the range $0.30-0.45$~mag
\citep{b94i1,rich94i} are reasonable and altering the reddening in
this range will not greatly affect our
results. Figure~\ref{apr15_calc2987} displays our best fit with our
massive model. Neither the CO21 model nor the massive model are able
to reproduce all of the observed features. In particular, neither of
the models predicts the strong observed absorption feature near
5800~\ang, which is likely produced by the Na D lines
\citep{millard94i99}. 

The CO21 model does a reasonable job at fitting
the overall lineshape of the observed infrared feature, which it does
with a blend of He~I and C~I $\lambda10695$ (and other C~I) as well as
O~I and Si~I-II lines. However, the CO21 model also produces strong,
unobserved optical absorption features near 5000~\ang, mostly due to
Fe~III. The CO21 model also does a rather poor job reproducing the
O~I~$\lambda7773$ feature and the Ca~II IR triplet. 

The massive model
has less of a problem with the strong absorption near 5000~\ang\ and
even begins to reproduce the observed Fe~II features in that region of
the spectrum, indicating that the ionization states of the iron-peak
elements is better reproduced in this model.  The O~I~$\lambda7773$
feature is even weaker in this model than in the CO21 model, but the
Ca~II IR triplet is better reproduced.  Both models do poorly at
reproducing the flat bottomed absorption feature near 6250~\ang, which
is produced either by the Si~II~$\lambda6355$ line or by the C~II
$\lambda6580$ line in a detached shell as proposed by
\cite{millard94i99}.  We discuss the identity of the features in detail
in \S~\ref{lineids}.

\begin{figure*}
\begin{center}
\leavevmode
\psfig{file=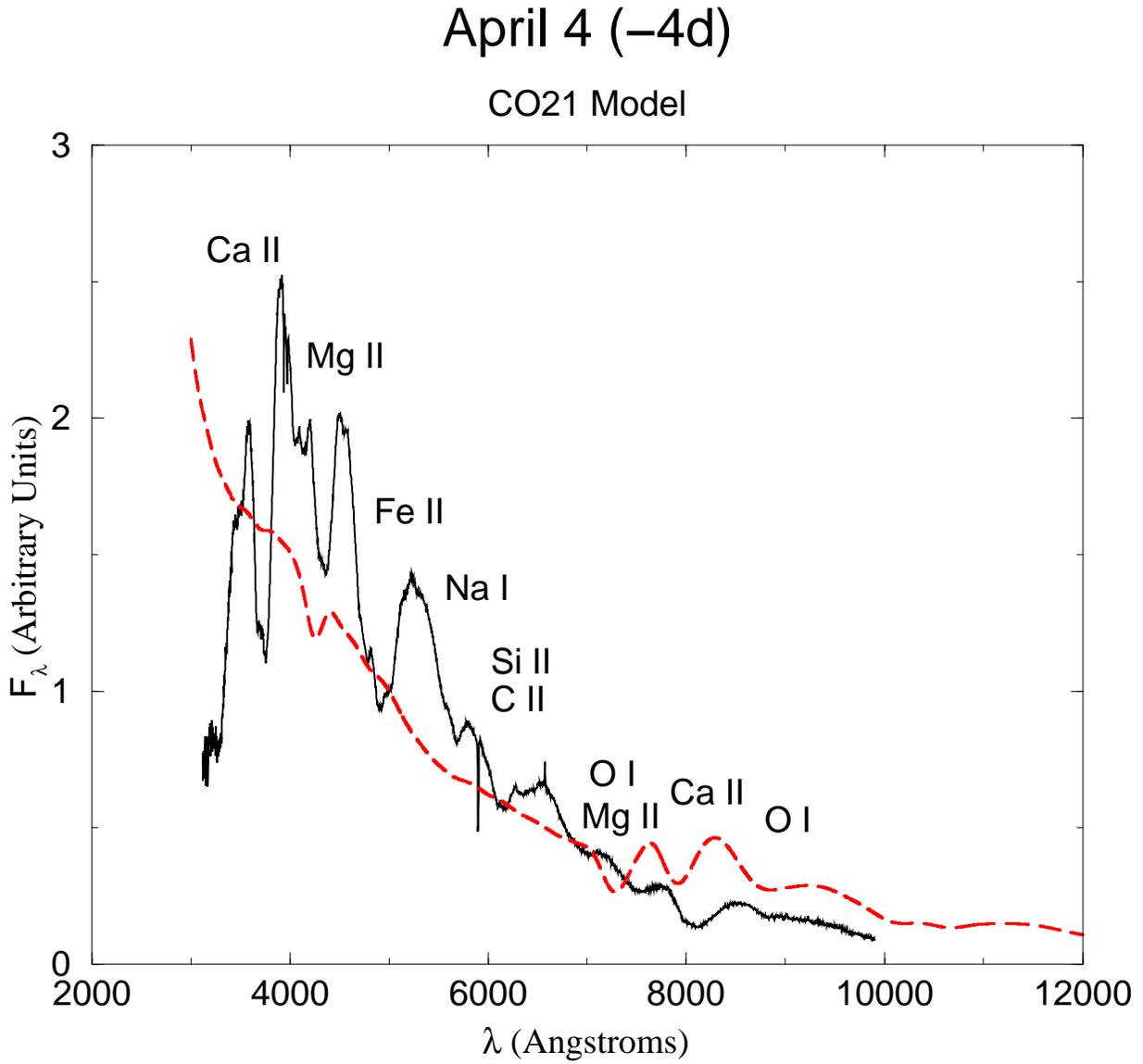}
\caption{\label{apr04_calc3560} The synthetic spectrum of the
\protect\cite{iwsn94i94} 
model (dashed curve) is compared to the  spectrum taken at Lick Observatory
on April 4, 1994 (-4d) \protect\citep{fil94i}.} 
\end{center}
\end{figure*}

\begin{figure*}
\begin{center}
\leavevmode
\psfig{file=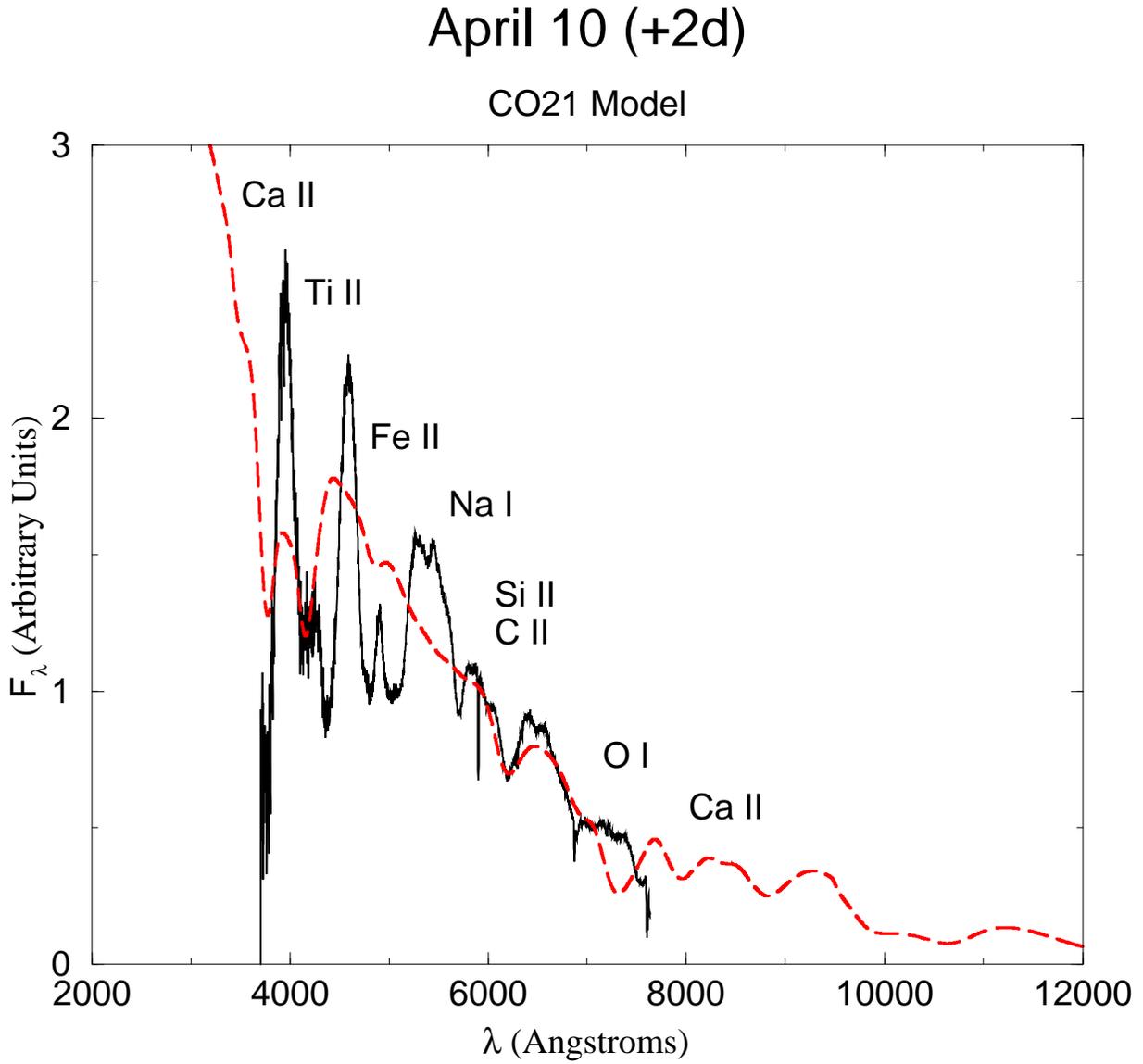}
\caption{\label{apr10_calc3432} The synthetic spectrum of the
\protect\cite{iwsn94i94} 
model (dashed curve) is compared to the  spectrum taken at the MMT
on April 10, 1994 (+2d) \protect\citep{skpc94i}.} 
\end{center}
\end{figure*}

\begin{figure*}
\begin{center}
\leavevmode
\psfig{file=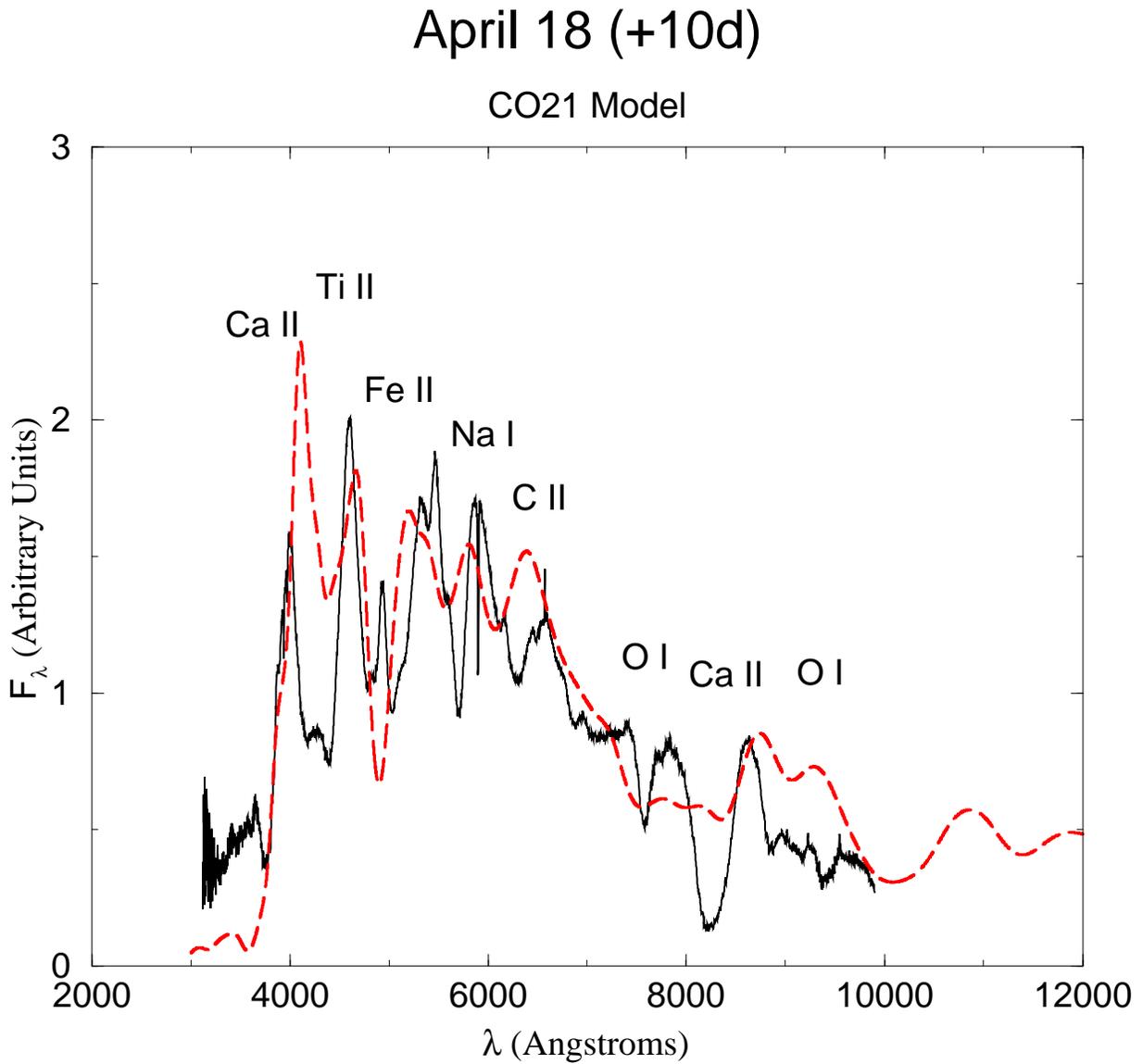}
\caption{\label{apr18_calc3454} The synthetic spectrum of the
\protect\cite{iwsn94i94} 
model (dashed curve) is compared to the  spectrum taken at Lick Observatory
on April 18, 1994 (+10d) \protect\citep{fil94i}.} 
\end{center}
\end{figure*}

Figures~\ref{apr04_calc3560}--\ref{apr18_calc3454} display the time
history of the CO21 model. In general the overall shape of the spectra
is reasonably well reproduced by the model, but the features
themselves are not. Figures~\ref{apr04_calc3560} shows that the model
calculation is  too blue at early times. This is almost certainly
due to the very low-mass in the ejecta and may also indicate that more
Fe is required in the models. Therefore the need for a somewhat more
massive model is 
indicated both by the early spectra and by the April 15 (+7d) spectrum.

\section{Discussion\label{lineids}}

\begin{figure*}
\begin{center}
\leavevmode
\psfig{file=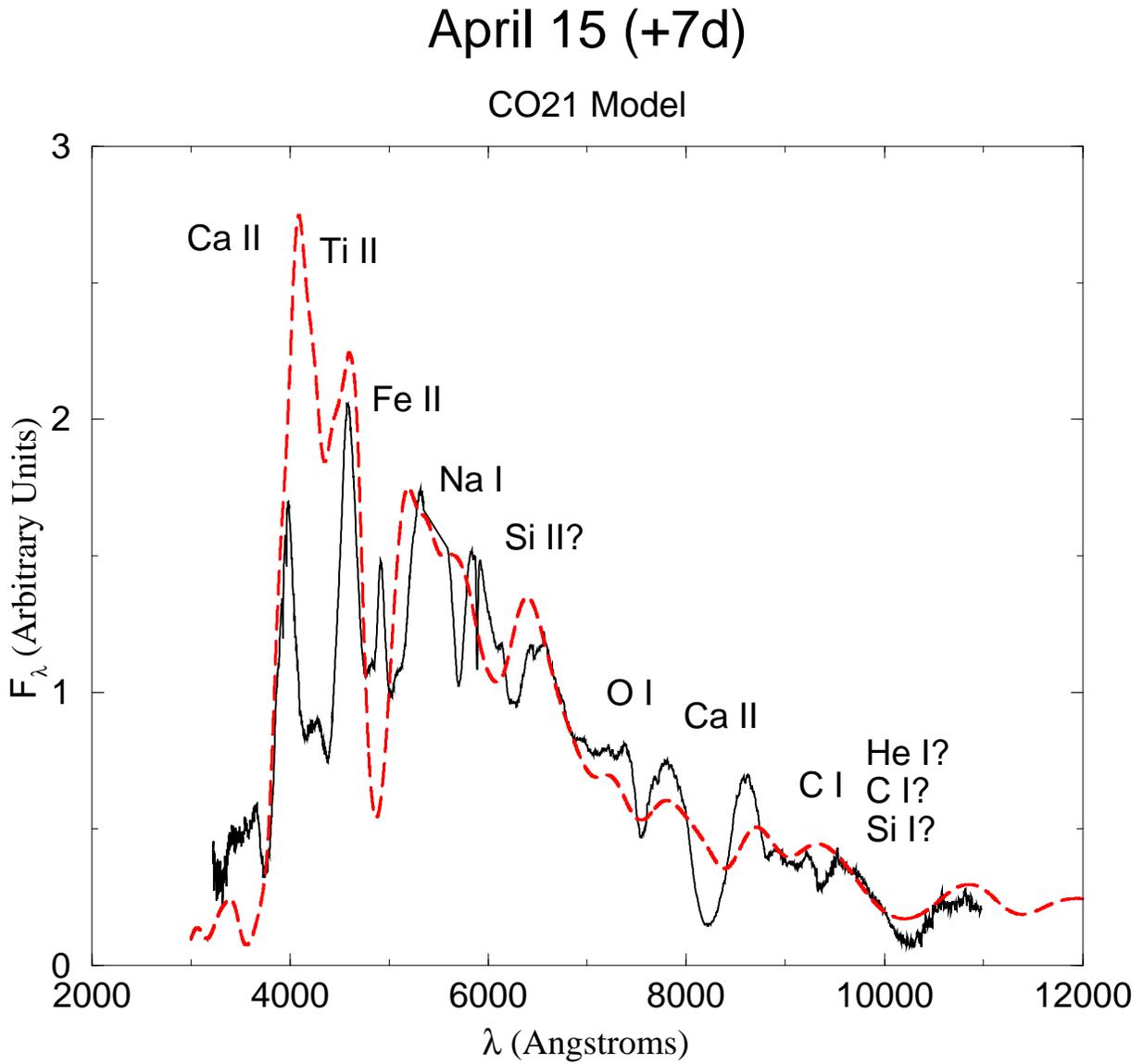}
\caption{\label{apr15_calc3463} The synthetic spectrum of the
\protect\cite{iwsn94i94} 
model (dashed curve) is compared to the  spectrum taken at Lick Observatory
on April 15, 1994 (+7d) \protect\citep{fil94i}. The line
identifications are the combined results of this work and that of
\protect\citet{millard94i99}.} 
\end{center}
\end{figure*}

\begin{figure*}
\begin{center}
\leavevmode
\psfig{file=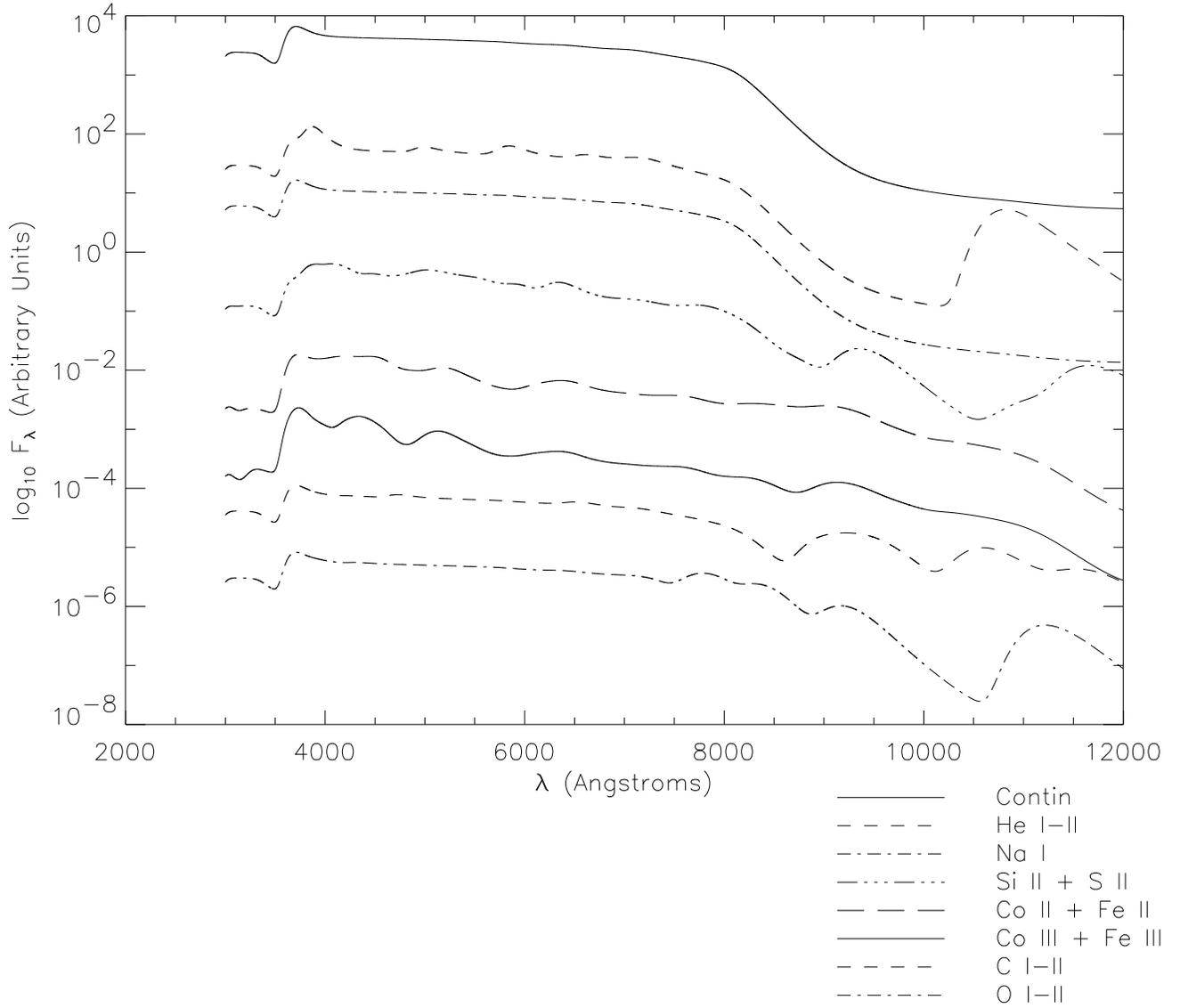}
\vspace*{60pt}
\caption{\label{line_ids}  The model spectra for the identical
conditions as displayed in Figure~\protect\ref{apr15_calc3463}, but
with only NLTE line opacity included; all LTE line opacity has been
set to zero. The legend indicates which particular NLTE lines are included
in the calculations.} 
\end{center}
\end{figure*}

Our models show that overall, an ejecta mass near 2.4~\msol, with a
mostly C+O+He composition, leads to reasonably good agreement with the
observed spectra near maximum light. Nevertheless, it is useful to
examine the deficiencies of the hydrodynamic model
and to explore what the possible line identifications in the observed
spectra are by using sophisticated models. This is especially important
once one realizes that in detailed calculations the features that are
usually identified as due to a single line are in reality produced by blends
of many different features that form in many places in the supernova
``atmosphere.'' Figure~\ref{line_ids} displays the spectra produced
using the structure of Figure~\ref{apr15_calc3463}, but with {\em all
LTE line opacities set to zero} and only the indicated NLTE line
opacities included; all continuum opacities are included in
all the calculations. Examining the spectrum with {\em only} continuum
opacities we see that the ``emission'' feature near 3800~\ang\ is in
fact the result of a (C~II) continuum edge. This fact should lead to
caution in identifying features (and deriving velocity information)
based solely on wavelength coincidences. The Fe~II and Co~II spectra lead 
to features that are recognizable and unremarkable although they
appear to be too weak in the synthetic spectrum. The Fe~III and Co~III
lines are too strong and lead to the unobserved features near 5000 and
6100~\ang, indicating that the iron-peak elements are
over-ionized. While the 6100~\ang\ feature is weak in this figure, we
believe it is masking the observed Na D feature. 
This is a clear sign that the mass of the CO21 model is too
low, since if the iron-peak elements were at higher density they would
be more likely to recombine.  The
temperature structure of the atmosphere,  (and therefore our
value of the total bolometric luminosity),  is reasonably well
determined 
since it is clearly required to reproduce the overall
shape of the spectrum. The He~I spectrum is
interesting, in that it produces a relatively weak He~I~$\lambda5876$
line and a He~I~$\lambda10830$ feature that is essentially in
emission. In the full spectrum (Figure~\ref{apr15_calc3463}) there is
no evidence of the He~I~$\lambda5876$ feature and clearly the
predicted
feature
due to He~I~$\lambda10830$ is too weak to have fully produced the
observed infrared feature; therefore in our models this feature must
be produced by 
 additional contributions 
from C~I, O~I, and Si~I--II (Si~I is treated in LTE in our
calculations). \citet{sfnw93} find that whenever He~I~$\lambda10830$
is present, so too are He~I~$\lambda5876$ and He~I~$\lambda6678$. While this
result lends credence to our suggestion that the IR feature is in fact
produced by alternate species, rather than by helium, we don't find
the He~I~$\lambda10830$ to be as pronounced as they do in their
calculations, which could be caused by the fact that we are examining
different epochs or that we treat more species and complex blending
effects diminish the strength of the He~I~$\lambda10830$ line. The spectrum
produced by Si~II in Figure~\ref{line_ids} 
is also notable in that it predicts a feature from the
Si~II~$\lambda6355$ line that is not noticeable in the full synthetic
spectrum.  Since the infrared feature is
produced to some extent in our calculations by Si~I, it is clearly
important to include this species in NLTE in future
work. \citet{millard94i99} have shown that in parameterized calculations
it is not possible to determine whether the feature ascribed to
He~I~$\lambda10830$ is due to a blend of He~I and C~I  or  to
a blend of 
Si~I lines. The fact that the infrared feature is produced by a blend
of a number of different species in our calculations should lead to
caution when making line identifications based solely on wavelength,
particularly when there is only a single feature from an identified
species.

\begin{figure*}
\begin{center}
\leavevmode
\psfig{file=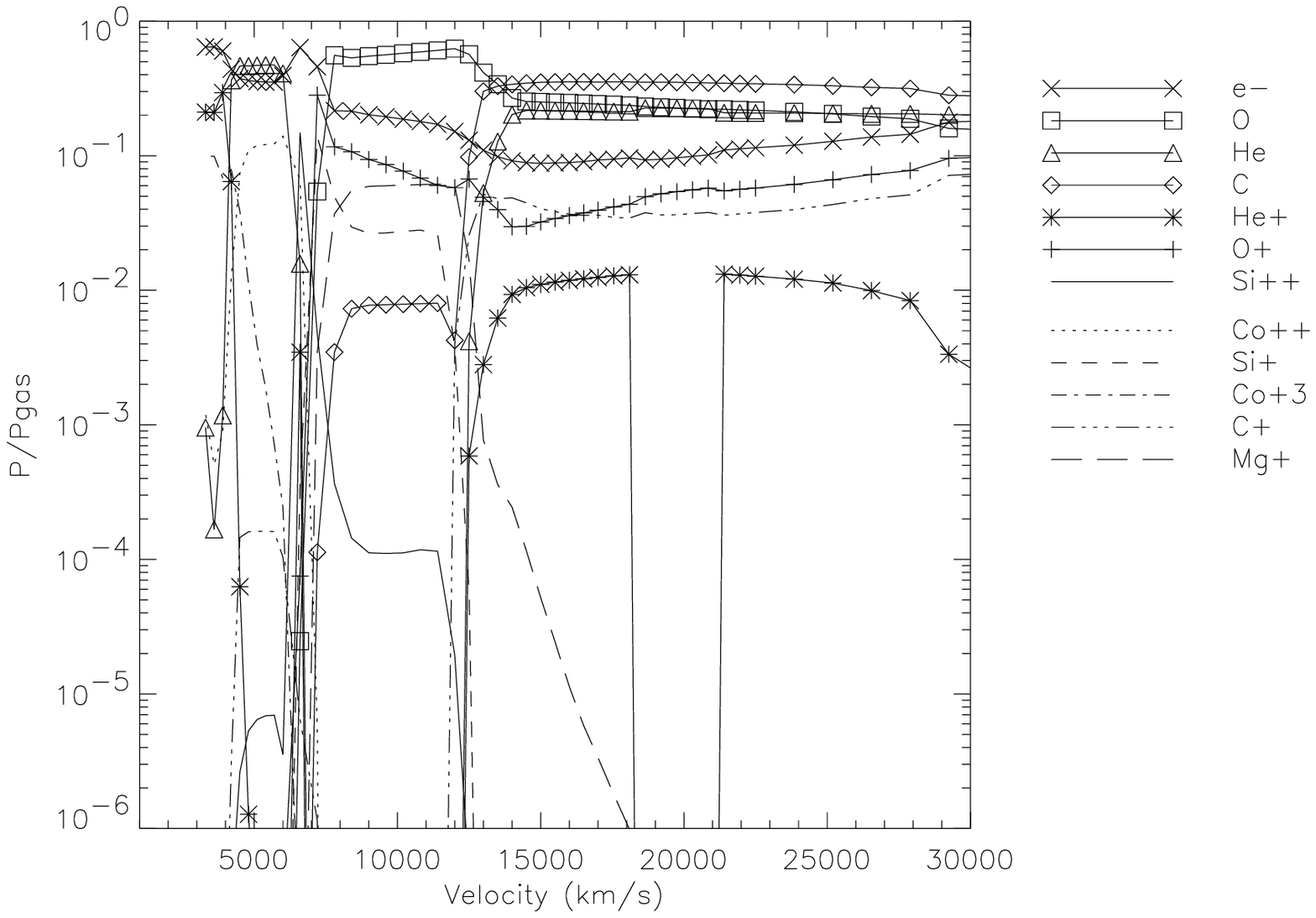}
\caption{\label{abundfid3_2987} 
The partial pressures of the most
prevalent species as a function of velocity in the parameterized more
massive model for the 
April~15 (+7d) epoch. (The synthetic spectrum produced by this model is shown in
Figure~\protect\ref{apr15_calc2987}.)}
\end{center}
\end{figure*}

\begin{figure*}
\begin{center}
\leavevmode
\psfig{file=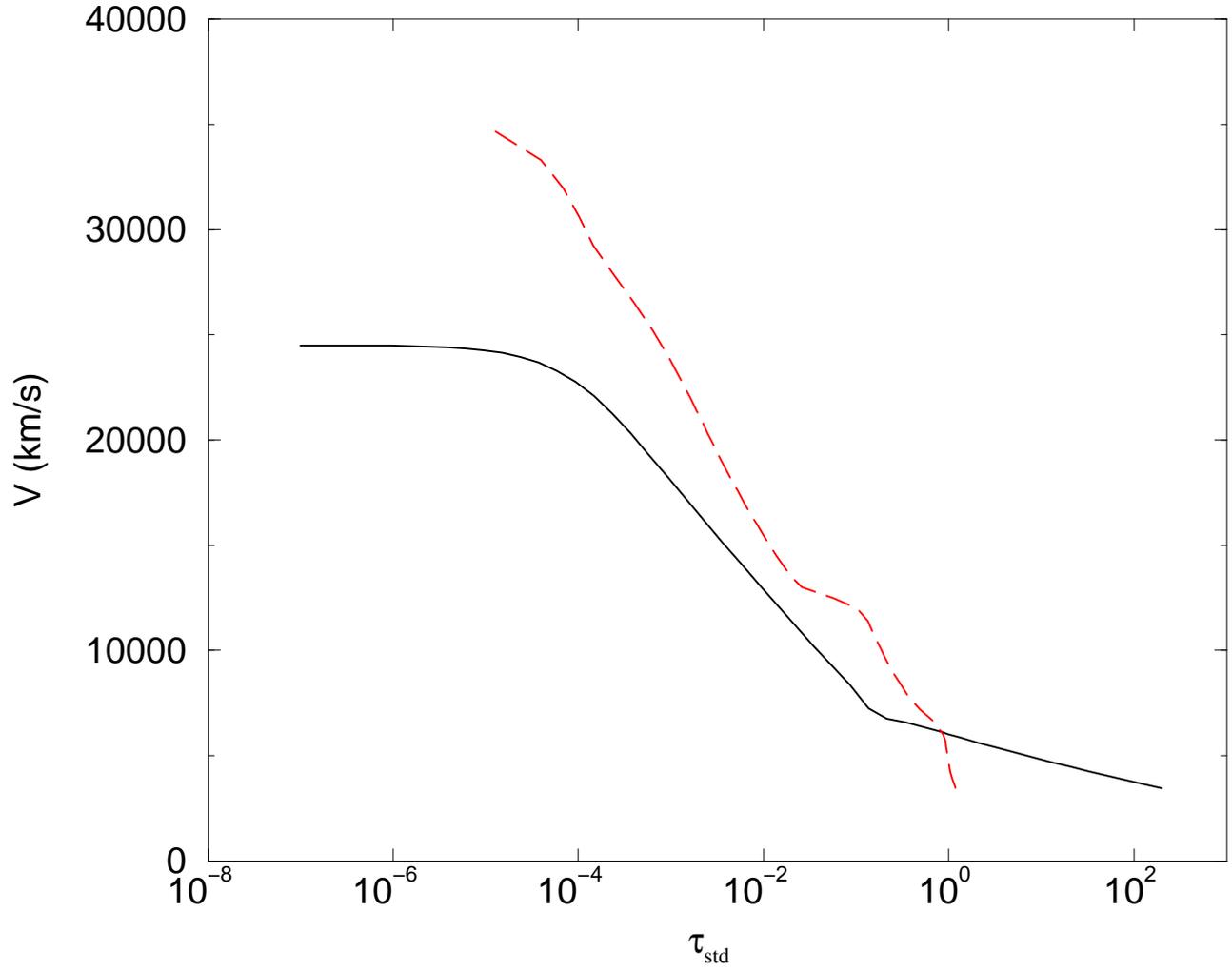}
\caption{\label{vels} 
The velocity profiles of the two models in
Figures~\protect\ref{abundco21_3463} (dashed line) and
\protect\ref{abundfid3_2987} (solid line). \tstd is the total
continuum optical depth at 5000 Angstroms.}  
\end{center}
\end{figure*}

Since the excited 
states of He~I lie so high above the ground state, a good predictor for
the strength of the He~I lines is the abundance of He~II. The fact
that He~II is prevalent in our parameterized model (see
Figure~\ref{abundfid3_2987}), and somewhat less 
prevalent in the CO21 model, probably leads to the more massive
parameterized model's better reproduction of the observed lineshapes
and lends some support for the identification of the line as due to helium.
This support should be tempered by the realization that our treatment
of gamma-ray deposition in the massive model is parameterized and
not based on direct hydrodynamical calculations. However, we have
varied the value of the deposition function by 4 orders of magnitude
which mimics a huge range in mixing (probably a larger range than is physically
reasonable). 
Additionally, the fact that in the CO21 model the observed
infrared feature  is produced by a blend of other
lines should lead to caution in the identification
of helium in SN~1994I in particular, and SNe~Ic in general. Although
\citet{clocc94i96} argue that the absence of a redward shift in the Na~D
feature is evidence for He~I $\lambda5876$ in SN 1994I, there is
another feature of the same strength just blueward that is
unidentified in their spectra; thus, while indicative, these features are not
conclusive evidence for the presence of He~I. A
hydrodynamical model that does a better job of reproducing the
spectrum is needed. Such a model would likely be more massive than
CO21, have the iron-peak elements concentrated in the first
ionization stage, and have more Si~II at low
velocity. Also, since the Na~D line is not well reproduced in any of
our models, we suspect that the gamma-ray deposition is
incorrectly ionizing  the sodium.

\citet{fassiaetal98} find that in order to fit the observed
infrared feature (which they assume to be due to He~I~$\lambda10830$)
in the Type II SN~1995V, they need both dredge-up 
of $^{56}$Ni into the hydrogen envelope as well as clumping of helium
within the hydrogen envelope. Given the fact that their models fail to
fit the O~I~$\lambda 7773$ line and that blending with O~I and
Si~I-II lines may be important in SNe~II as well as SNe~Ib/c, it seems
prudent to view any conclusions based solely on the apparent presence of
He~I~$\lambda10830$  in SNe with  caution. Clearly detailed
synthetic spectral models of high-quality hydrodynamic simulations is
the best way to draw firm conclusions about the structure of the
supernova ejecta.

We note that since we have not calculated the spectra produced by the
model of Woosley and collaborators \citep{wlw93,wooseastasi97}, our
results certainly are compatible with that model and it should still
be considered a viable candidate. Our most important result is,
however, that the IR feature that has been attributed to helium, is
\emph{not} a clear indicator of the presence of helium, since it can
be produced both by a blend of Si~I lines \citep{millard94i99}, or at
the very least consists of a blend of helium and carbon. All the
models that we have explored where the feature was produced explicitly
by helium showed strong optical helium features in agreement with the
results of \citet{millard94i99}. Finally, we note that future
explosion models should  
focus on  more massive progenitors and 
should consider models with and without helium in order to
determine the identity of the observed IR-feature. 
 
\acknowledgments We thank Peter H{\"o}flich and Ken Nomoto for helpful
discussions and for providing us with the CO21 model.  This work was supported
in part by NSF grants AST-9417213, AST-9417242, AST-9731450, and
AST-9417102; NASA 
grant NAG5-3505; an IBM SUR grant to the University of Oklahoma; and
NSF grant AST-9720704, NASA ATP grant
NAG 5-3018, and LTSA grant NAG 5-3619 to the University of Georgia.
Some of the calculations presented in this paper were
performed at  the San Diego
Supercomputer Center (SDSC), supported by the NSF, and at the National
Energy Research Supercomputer Center (NERSC), supported by the
U.S. DOE. We thank both of these institutions for a generous allocation
of computer time.

\bibliography{refs,sn1bc,sn1a,snii,gals,crossrefs}


\begin{deluxetable}{rrrr}
\tablecolumns{4}
\tablewidth{0pc}
\tablecaption{\label{nltetab}NLTE Species}
\tablehead{\colhead{Element} &  \multicolumn{3}{c}{Ionization Stage}}
\startdata
        &   I  & II & III \\
\hline
H       &  15/105& \nodata & \nodata \\
He      &  11/14 & \nodata & \nodata \\
C       &  228/1387 & 85/336 & 79/365  \\
O       &  36/66 & 171/1304 & \nodata \\
Ne      &  26/37 & \nodata & \nodata \\
Na      &  3/2 & \nodata & \nodata \\
Mg      &\nodata  &  18/37 & \nodata \\
Si      &\nodata  & 93/436 & 155/1027  \\
S       &\nodata  & 84/444 & 41/170  \\
Ca      &\nodata  & 87/455 & \nodata \\
Fe      & 494/6903 & 617/13675&566/9721 \\
\tablecomments{The species and number of
levels/transitions treated in non--LTE by {\tt PHOENIX} in these
calculations.}
\enddata
\end{deluxetable}

\clearpage

\clearpage

\clearpage

\clearpage

\clearpage

\clearpage

\clearpage

\clearpage

\clearpage

\clearpage

\end{document}